\documentclass[sn-mathphys]{sn-jnl}

\jyear{2021}%

\theoremstyle{thmstyleone}%
\theoremstyle{thmstyletwo}%

\theoremstyle{thmstylethree}%

\newcommand{\dd}{{\rm{d}}}                                 
\newcommand{\U}{{\mathcal{U}}}
\newcommand{\V}{{\mathcal{V}}}
\newcommand{\h}{{\mathcal{H}}}
\newcommand{\up}{u_+}
\newcommand{\Up}{{\mathcal U}_+}
\newcommand{\lp}{\lambda_+}
\newcommand{\G}{{\mathcal{G}}}
\newcommand{\imag}{{\mathrm{i}}}

\begin{document}

\title[Penrose junction conditions with $\Lambda$]
{Penrose junction conditions with $\Lambda$:\\
Geometric insights into low-regularity metrics \\
for impulsive gravitational waves}



\author*[1]{\fnm{Ji\v{r}\'{\i}} \sur{Podolsk\'y}}\email{jiri.podolsky@mff.cuni.cz}

\author[2]{\fnm{Roland} \sur{Steinbauer}}\email{roland.steinbauer@univie.ac.at}

\affil[1]{\orgdiv{Institute of Theoretical Physics}, \orgname{Charles University, Faculty of Mathematics and Physics}, \orgaddress{\street{V Hole\v{s}ovi\v{c}k\'ach 2}, \city{Prague 8}, \postcode{18000}, \country{Czechia}}}

\affil[2]{\orgdiv{Faculty of Mathematics}, \orgname{University of Vienna}, \orgaddress{\street{Oskar-Morgenstern-Platz 1}, \city{Vienna}, \postcode{1090}, \country{Austria}}}


\abstract{
Impulsive gravitational waves in Minkowski space were introduced by Roger Penrose at the end of the 1960s, and have been widely studied  over the decades. Here we focus on nonexpanding waves which later have been generalized to impulses traveling in all constant-curvature backgrounds, i.e., the (anti-)de Sitter universe. While Penrose's original construction was based on his vivid geometric ``scissors-and-paste'' approach in a flat background, until recently a comparably powerful visualization and understanding has been missing in the case with a cosmological constant ${\Lambda\not=0}$. Here we review the original Penrose construction and its generalization to non-vanishing $\Lambda$ in a pedagogical way, as well as the recently established visualization: A special family of global null geodesics defines an appropriate comoving coordinate system that allows to relate the distributional to the continuous form of the metric.
}




\maketitle

\section{Introduction}\label{sec1}

In this paper we would like to pay tribute to Sir Roger Penrose for his lifelong contribution to mathematical physics and to general relativity, in particular. The list of fundamental ideas, concepts and methods he has shaped is incredibly vast, ranging from his celebrated singularity theorem and his cosmic censorship hypotheses to the Newman--Penrose formalism and twistor theory, to name only a few. Here we wish to review yet another topic in mathematical relativity he has pioneered and which is still an active area, namely \emph{impulsive gravitational waves}.

It was at the end of the 1960s when Roger Penrose introduced this topic in \cite{Penrose:1968, Penrose:1968a}. The work \cite{Penrose:1968a} actually is a written version of a lecture series delivered at the Battelle Seattle Research Center in the summer of 1967 on differential geometry, spinors and spacetime singularities. Impulsive plane waves appear there on page~198 as an example of a spacetime which does not possess a Cauchy surface, simplifying an earlier example of an extended plane wave given in \cite{Pen:65}, which exploits the focusing effect the wave exerts on null geodesics. Such impulsive waves are introduced as idealized versions of sandwich waves with infinitesimal duration but still producing an effect in the sense that the wave profile is a Dirac-delta. Since such metrics clearly do not satisfy the usual regularity assumptions for spacetimes which possess a delta-function curvature on a hypersurface,\footnote{These assumptions put on the metric are that it is $C^0$ everywhere, but fails to be $C^3$ on a hypersurface.} Penrose also described a geometric construction using a vivid visualization that leads to a continuous metric, which models the same situation. This construction was more explicitly given in \cite[p.~82f.]{Penrose:1968} where also the term ``scissors-and-paste'' occurs for the first time. The focus of this work, however, was to employ impulsive \emph{pp}-waves as an example illustrating the construction of spacetime twistors.

Finally,  Penrose's seminal paper \cite{Penrose:1972}, which was a contribution to the volume in honour of J.~L.~Synge, was entirely devoted to the geometry of impulsive waves in Minkowski space. It is here that the continuous metric is for the first time given explicitly (in the plane wave case), and that also spherical impulsive waves are considered. Again, the geometry of the (single) null wave surface is studied  using spinors.

From there on, impulsive gravitational waves have been used in many contexts as models of short but violent bursts of gravitational radiation. Over the years they have attracted the attention of researchers in exact spacetimes, who have widely generalized the original class of solutions, of particle physicists, who have used them as toy models in quantum scattering, and of geometers, who have used them as relevant key-models in low regularity Lorentzian geometry.


Personally, the above mentioned works of Penrose have been a source of inspiration for us during many years. It is thus an honour for us to review Penrose's geometric constructions and some of its generalizations in this contribution. In particular, we will concentrate on \emph{nonexpanding} impulsive gravitational waves in (anti-)de Sitter space, and put the respective geometric constructions in the context of low-regularity Lorentzian geometry.

More precisely, we will recall Penrose's ingenious ``scissors-and-paste'' construction (nowadays and in the following called ``cut-and-paste'' method) of impulsive waves in flat space in Section~\ref{sec2}. Then, in Section~\ref{sec3} we will briefly discuss the distributional as well as the continuous metric forms for impulsive \emph{pp}-waves in Minkowski space \cite{PodolskyVesely:1998}, and also their interrelation. In fact, we sketch the consistent mathematical way of \cite{KunzingerSteinbauer:1999b} looking at the ``discontinuous coordinate transform'' between them. In Section~\ref{sec4} we move on to explain the generalization of the Penrose construction to nonexpanding impulsive waves in (anti-)de Sitter space \cite{PodolskyGriffiths:1999} and explicitly derive, again, the distributional as well as the continuous form of the metric. In Section~\ref{sec5} we turn to discussing the interrelation between these two metric forms by studying a special family of null geodesics crossing the impulse. These recent calculations \cite{PodolskySaemanSteinbauerSvarc:2019} finally lead to a geometric and vivid picture which we will present in Section~\ref{sec6}, generalizing the original visualization of Penrose to the case ${\Lambda\not=0}$.

\section{Penrose's construction of plane and spherical impulsive waves in Minkowski space}
\label{sec2}

In this section we recall the  beautiful geometric construction of impulsive waves propagating in flat space given by Roger Penrose in \cite{Penrose:1968, Penrose:1968a} and, most importantly, in \cite{Penrose:1972}. The basic idea is the following:

Minkowski space is ``cut'' along a \emph{null plane} into two ``halves'',  which are then re-attached with a ``warp'', given by the so-called \emph{Penrose junction conditions}. This ``cut-and-paste'' approach leads to the construction of an impulsive \emph{pp}-wave. In particular, Penrose considered a specific \emph{plane wave}.

In the same work \cite{Penrose:1972}, Penrose also constructed impulsive \emph{spherical wave} as a single sphere of curvature that expands at the speed of light. In this case, Minkowski space is cut along a \emph{null cone}, and the junction conditions are more involved.
\vspace*{2mm}

In both cases, the {\bf Penrose geometric recipe} for the construction of an impulsive wave in Minkowski space ${\cal M}$ is the following:

\begin{itemize}

\item {\bf cut} the space ${\cal M}$ along the null plane or null cone ${\cal N}$ using ``scissors'',

\item {\bf shift} the two resulting half-spaces ${\cal M}^-$, ${\cal M}^+$ along the cut with a ``warp'',\footnote{Here we re-attach ${\cal N}$ to both ${\cal M}^-$ and ${\cal M}^+$, and consider them as manifolds with boundary ${\cal N}$.}

\item {\bf paste} them together identifying the corresponding boundary points in ${\cal N}$.

\end{itemize}

Let us now be more specific and present this construction explicitly using the most natural coordinates for such a procedure. We start with the plane wave case.

\subsection{Plane impulsive waves}
\label{subsec2.1}

\begin{figure}[h]%
\centering
\includegraphics[scale=0.50]{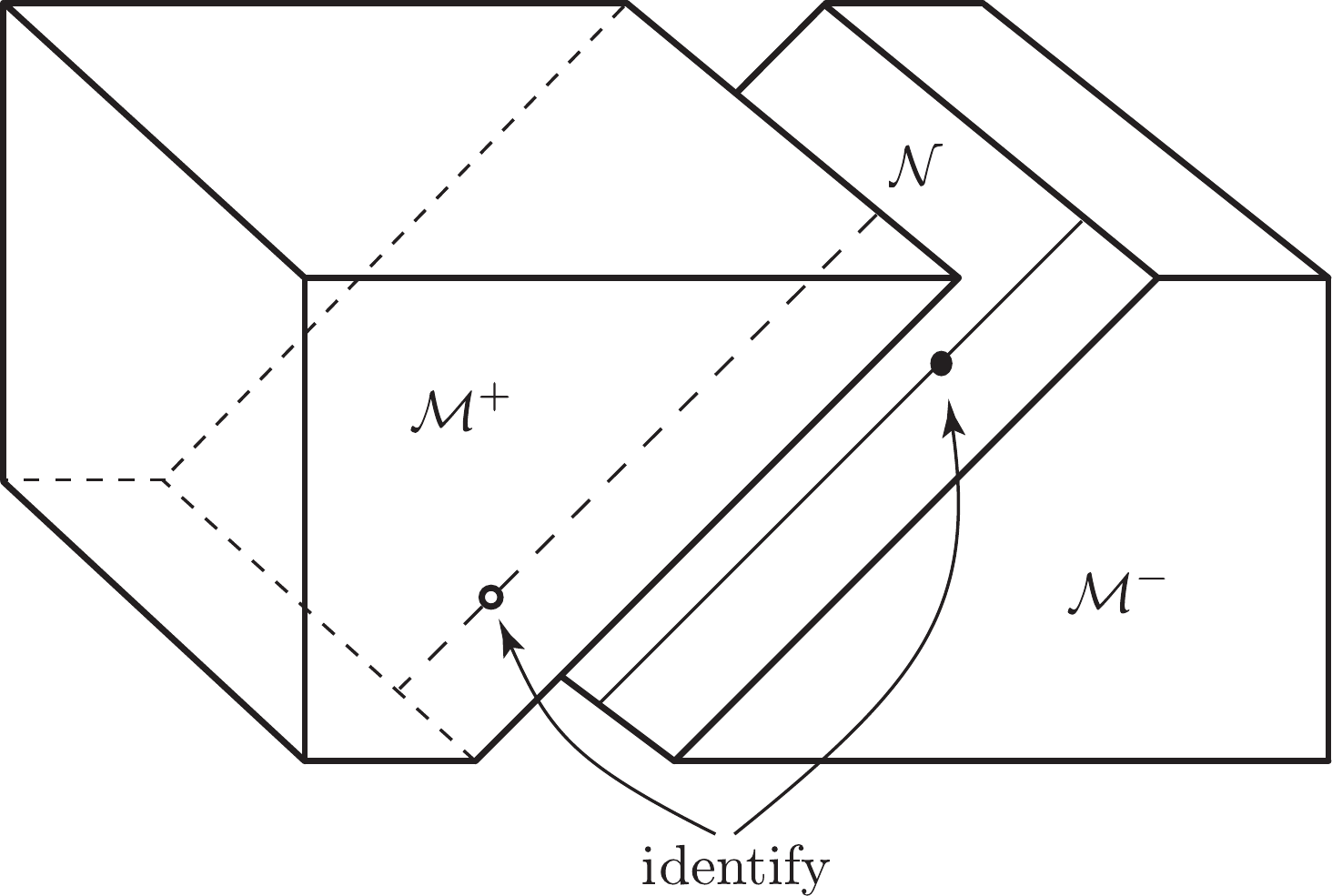}
\caption{Minkowski space is cut into two parts ${\cal M}^-$ and ${\cal M}^+$ along a null plane ${\cal N}$. These parts are then re-attached with a ``warp'' in which points are arbitrarily shunted along the null generators of the cut, and then identified. Such a construction generates an impulsive plane gravitational wave.}
\label{fig1}
\end{figure}

In this case the usual \emph{null coordinates} of flat space are employed, namely
\begin{equation}
\U = \tfrac{1}{\sqrt2}\,(t-z)\,, \qquad   \V = \tfrac{1}{\sqrt2}\,(t+z)\,, \qquad
\eta = \tfrac{1}{\sqrt2}\,(x+\imag\,y)\,, \label{Mink-null-coordinates}
\end{equation}
in which the Minkowski metric takes the form
\begin{equation}\label{backgr-Mink}
  \dd s_0^2 = 2\,\dd\eta\,\dd\bar\eta -2 \,\dd\U\,\dd\V\,.
\end{equation}
Now this spacetime ${\cal M}$ is cut along the {\bf null plane} ${\cal N}$ given by ${\U=0}$, and the half-spaces ${\cal M}^-$ and ${\cal M}^+$ are defined by ${\U<0}$ and ${\U>0}$, respectively, see Figure~\ref{fig1}.

\vspace{-1mm}

The warp at ${\U=0}$ is then given by a \emph{deformed shift along} $\V$, specified by an arbitrary function $\h(\eta,\bar\eta)$, while keeping $\eta$ (that is $x,y$) fixed. These {\bf Penrose junction conditions} at ${\U=0}$ are explicitly given by
\begin{equation}
\mbox{\fbox{
{$\big(\V,\eta,\bar\eta\big)_{{\mathcal M}^-}\,=\,\big(\V-\h(\eta,\bar\eta),\eta,\bar\eta\big)_{{\mathcal M}^+}\,. $}
}}
\label{Penrose-junction-conditions}
\end{equation}
Such an identification of the boundary points leads to an impulsive plane wave characterized by the (arbitrary) real function $\h(\eta,\bar\eta)$, as we will see more explicitly in Section \ref{sec3}.

\subsection{Spherical impulsive waves}
\label{subsec2.2}

\begin{figure}[h]%
\centering
\includegraphics[scale=0.63]{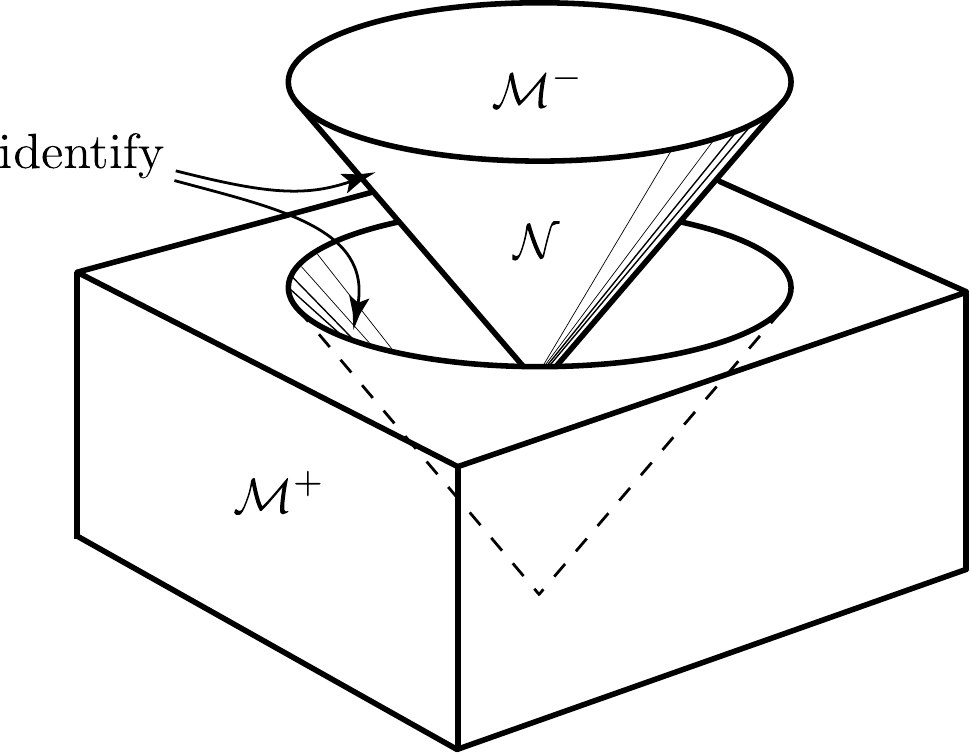}
\caption{Minkowski space is cut into two parts ${\cal M}^-$ and ${\cal M}^+$ along a future null cone ${\cal N}$. The parts are then re-attached with an arbitrary ``warp'' by identifying points on the two boundaries. Such a construction generates an impulsive spherical wave.}
\label{fig2}
\end{figure}

To obtain spheres expanding at the speed of light, alternative coordinates of Minkowski space \eqref{backgr-Mink}, namely
\begin{equation}
\U = U+ \zeta\bar\zeta\, V\,, \qquad   \V = V\,, \qquad
\eta = \zeta\,V\,, \label{Mink-null-cone-coordinates}
\end{equation}
must be employed in which ${U=0}$ represents a {\bf null~cone}~${\cal N}$. In fact, these are the Robinson--Trautman coordinates
\begin{equation}\label{backgr-spher}
  \dd s_0^2 = 2\, V^2\,\dd\zeta\,\dd\bar\zeta -2 \,\dd U\,\dd V \,,
\end{equation}
where ${V=0}$ labels the vertex of the cone, from which the spherical impulse expands (these coordinates degenerate along ${t=-z}$).
The ``half-spaces'' ${\cal M}^-$ and ${\cal M}^+$, again given by ${U<0}$ and ${U>0}$, are now the interior and the exterior of the null cone, respectively, see Figure~\ref{fig2}.

\vspace{-1mm}

The warp at ${U=0}$ is performed by the {\bf Penrose junction conditions}
\begin{equation}
\mbox{\fbox{
{$\displaystyle\big(V,\zeta,\bar\zeta\big)_{{\mathcal M}^-}\,=\,\Big(\frac{V}{\lvert{{h}'}\lvert},
h(\zeta),\bar h(\bar\zeta)\Big)_{{\mathcal M}^+}\,, $}
}}
\end{equation}
where $h(\zeta)$ is an arbitrary holomorphic function of the complex coordinate $\zeta$ (which is actually a stereographic representation of the spherical angles $\theta, \phi$ on the expanding impulse). Such an identification of the boundary points represented by the mapping ${\zeta \mapsto h(\zeta)}$ leads to an impulsive spherical wave, whose specific character is determined by $h(\zeta)$.

More details and many references can be found in Chapter 20 of \cite{GriffithsPodolsky:2009}, in the review \cite{Podolsky:2002}, or in \cite{BH:03}. Recent summaries of nonexpanding impulsive waves are contained in \cite{PodolskySaemannSteinbauerSvarc:2015}, and of expanding (spherical) impulsive waves in~\cite{PodolskySaemannSteinbauerSvarc:2016}.

Our present contribution concentrates on \emph{nonexpanding} impulses propagating in Minkowski, de Sitter and anti-de Sitter spaces (maximally symmetric vacuum spacetimes with any value of the cosmological constant $\Lambda$), and is based mainly on our recent papers on this topic \cite{SaemannSteinbauerLeckePodolsky:2016,SaemannSteinbauer:2017}, and most of all \cite{PodolskySaemanSteinbauerSvarc:2019}.

\section{Impulsive pp-waves}
\label{sec3}

In this section we wish to discuss \emph{planar} impulses in flat \emph{Minkowski} space. In fact, these impulsive plane-fronted waves with parallel rays geometrically belong to the famous family of \emph{pp-waves} \cite{SKMHH:2003, GriffithsPodolsky:2009}.

\subsection{Continuous and distributional metric forms}
\label{subsec3.1}

In his work \cite{Penrose:1972} Roger Penrose not only introduced the geometrical ``cut-and-paste'' construction method (described in Section~\ref{subsec2.1}) but also presented both a \emph{continuous} and a \emph{distributional} metric form of \emph{impulsive pp-waves}, and their mutual relation.

While in \cite{Penrose:1972} only a particular warping function was considered explicitly, namely the quadratic expression ${\h = \eta^2+\bar\eta^2}$ (which enters the metric \eqref{distrib-0} below), the procedure also works for the complete family of \emph{pp}-waves parametrized by an arbitrary function ${\h(\eta,\bar\eta)}$. Indeed, extending Penrose's original idea we may apply to the flat metric \eqref{backgr-Mink}, that is  to ${\,\dd s_0^2 = 2\,\dd\eta\,\dd\bar\eta -2 \,\dd\U\,\dd\V\,}$, the transformation

\vspace{-1mm}
\begin{equation}
\fbox{\begin{minipage}{5.6cm}
\vspace{-0.5cm}
  \begin{eqnarray}
  \  {\U} &=& u \,, \nonumber \\
  \  {\V} &=& v+ \Theta(u)\,h+\up(u)\,h_{,Z}\,h_{,\bar Z} \,,\nonumber\\
  \  \eta &=& Z+\up(u)\,h_{,\bar Z} \,,\nonumber
  \end{eqnarray}
\vspace{-0.7cm}
\end{minipage}}
\label{trans}
\end{equation}
where
\begin{equation}\label{h}
h\equiv h(Z,\bar Z)
\vspace{0.2cm}
\end{equation}
is \emph{any smooth enough real-valued function} of the complex variable $Z$ and its complex conjugate $\bar Z$. Moreover, $\up$ is the \emph{(Lipschitz) continuous kink function}, while $\Theta$ is the (locally bounded, i.e., $L^\infty_{\mbox{{\tiny loc}}}$) \emph{Heaviside step function}
\begin{equation}\label{kink-Heaviside}
   \up(u)=  \begin{cases}
       \ 0 &\mbox{for $u \le 0$}\\
       \ u &\mbox{for $u \ge 0$}
        \end{cases}
  \,, \qquad\qquad
  \Theta(u)=  \begin{cases}
       \ 0 &\mbox{for $u < 0$}\\
       \ 1 &\mbox{for $u > 0$}
        \end{cases}
  \,.
\end{equation}
Therefore, ${h(Z,\bar Z)=h(\eta,\bar \eta)}$ at ${u=0}$, which may be identified with $\h(\eta,\bar\eta)$.

If \eqref{trans} is applied \emph{separately} to \eqref{backgr-Mink} for ${u<0}$ defining ${\mathcal M}^-$ (where it is just an identity ${{\U}=u, {\V}=v, \eta=Z}$) and to ${u>0}$ defining ${\mathcal M}^+$, the metric becomes \begin{equation}
\mbox{\fbox{
{
$\dd s^2 = 2 \left\lvert\dd Z+\up(u)(h_{,\bar Z Z}\,\dd Z+{h}_{,\bar Z\bar Z}\,\dd\bar Z) \right\lvert^2
  -2\,\dd u\,\dd v\,.$
}
}}\label{conti-0}
\end{equation}
\nopagebreak[3]
This is the continuous {\bf Rosen form} of a \emph{pp}-wave \cite{AB:97,PodolskyVesely:1998}, which is impulsive due to the {\bf (mere) Lipschitz continuity} of the coefficient $\up$.\footnote{Rosen presented this type of the metric only for (extended) plane waves, which are a special subcase of the complete family of {\it pp}-waves.}


Interestingly, applying the transformation \eqref{trans} to  \eqref{backgr-Mink} for \emph{any} ${u\in \mathbb{R}}$ we \emph{formally} get the metric with the Dirac delta $\delta({\U})$,\footnote{More precisely, by applying the discontinuous transformation \eqref{trans} on the metric \eqref{distrib-0}, with the distributional identities ${\up'=\Theta}$, ${\Theta'=\delta}$, ${\Theta\up=\up}$ and the identification ${\h\lvert_{u=0}\,=h}$ (using ${\eta\lvert_{u=0}\,=Z}$), one obtains the continuous Rosen form \eqref{conti-0}.}
\begin{equation}
\mbox{\fbox{
{\
$\dd s^2 = 2\,\dd\eta\,\dd\bar\eta-2\,\dd \U\,\dd \V+2\,\h(\eta,\bar\eta)\,\delta({\U})\,\dd {\U}^2\,.$
}
}}\label{distrib-0}
\end{equation}
This is the {\bf Brinkmann form} of a \emph{pp}-wave, which is impulsive due to it being explicitly {\bf distributional}~in~$\,\U\,$. Its warping metric function is given by ${\,\h(\eta,\bar \eta)\equiv h(Z,\bar Z)\,}$ evaluated at ${\,u=0\,}$.

\vspace*{2mm}

Moreover, it can be immediately observed that the transformation \eqref{trans} is {\bf discontinuous} due to the presence of the \emph{Heaviside function $\Theta(u)$} {\bf entering} $\V$, which {\bf exactly represents the Penrose junction conditions} \eqref{Penrose-junction-conditions}, namely
${\big(\V,\eta,\bar\eta\big)_{{\mathcal M}^-}\,=\,\big(\V-\h(\eta,\bar\eta),\eta,\bar\eta\big)_{{\mathcal M}^+}}$. Recall that there is no change in $\eta$ at ${\U\equiv u=0}$ because $\eta(Z, \bar Z)$ given by expression \eqref{trans} is continuous.

\vspace{2mm}

There are thus \emph{close relations} between the continuous Rosen metric form~\eqref{conti-0}, the distributional Brinkmann metric form \eqref{distrib-0}, and the Penrose junction conditions \eqref{Penrose-junction-conditions} for impulsive \emph{pp}-waves. However, at this stage, these relations have to be considered only formal, because they involve distributions and also their products. A more rigorous treatment of the related mathematical subtleties occurring in low regularity is thus required to clarify exact meaning of these relations.

\subsection{Rigorously relating the continuous and distributional metric forms}
\label{subsec3.2}

To begin with, we discuss the regularities of the involved metrics. The continuous form \eqref{conti-0} of the impulsive \emph{pp}-wave is actually \emph{locally Lipschitz continuous}, a class of metrics which is often denoted by $C^{0,1}$ or $C^{1-}$. Such metrics are well within the Geroch--Traschen (or GT) class of metrics \cite{GT:87} which possess regularity ${H^2_{\mbox{{\tiny loc}}}\cap L^\infty_{\mbox{{\tiny loc}}}}$ and are uniformly nondegenerate \cite{LM:07,SV:09}. In their classical paper \cite{GT:87} Robert Geroch and Jennie Trashen  have shown that such metrics allow to (stably) define the Riemann tensor as a tensor distribution, and that they are well-suited to describe spacetimes which possess a \emph{distributional curvature supported on a hypersurface}. This is in fact the case for the metric \eqref{conti-0} which has the Riemann and the Ricci tensor proportional to $\delta(u)$, and hence its \emph{curvature concentrated on the impulse}.

On the other hand, the distributional form of the impulsive \emph{pp}-wave metric \eqref{distrib-0} clearly is \emph{outside} the GT-class, and hence \emph{there is no consistent distributional framework} (such as \cite{Mar:68}) available to study its curvature. Nevertheless, using the special Brinkmann coordinates it is formally possible to compute the curvature which then is again concentrated on the impulse and proportional to $\delta(\U)$. As a warning it has to be remarked that we have definitely reached the ``grey areas'' of distribution theory since e.g. only the mixed components ${R^i}_j$ of the Ricci tensor can be computed, but not those with both upper or lower indices. Moreover, the discontinuous change of coordinates is literally non-sensical within distribution theory since it boils down to performing the distributional pullback of the metric \eqref{distrib-0} by a merely $L^\infty$-map.

\vspace{2mm}

A {\bf rigorous investigation} of impulsive \emph{pp}-waves was performed in 1998--1999 by Michael Kunzinger and the second author in the series of articles \cite{Steinbauer:1998,KunzingerSteinbauer:1999a,KunzingerSteinbauer:1999b}.
First, in \cite{Steinbauer:1998} the geodesics (and the geodesic deviation) for the distributional form of impulsive \emph{pp}-waves were studied using a careful regularization procedure.
Thereby the Dirac $\delta$ in \eqref{distrib-0} was replaced by a \emph{general class of smooth functions}, the so-called {\bf model delta nets} $\delta_\varepsilon$ defined as follows: Choose a smooth function $\rho$ with unit integral, supported in $[-1,1]$, and set $\delta_\varepsilon(x)=(1/\varepsilon)\,\rho(x/\varepsilon)$.\footnote{All results are actually independent of the specific choice of $\rho$.} Technically, the geodesic
equation for the regularized metric(s) become nonlinear, and the fact that (at least for small regularization parameters $\varepsilon$) the geodesics exist long enough to cross the regularised impulse\footnote{By this we mean the support of the sandwich profile $\delta_\varepsilon$, i.e., the region $-\varepsilon\leq u\leq\varepsilon$.} and hence are complete, is established using a fixed point argument. The resulting geodesics for the distributional metric (obtained by a distributional limit) are then independent of the specific regularisation used, and reproduce earlier ad-hoc results of e.g.\ \cite{FPV:88,Bal:97}.

Then in  \cite{KunzingerSteinbauer:1999a} this analysis was put into the framework of {\bf nonlinear distributional geometry} \cite[Ch.~3.2]{GKOS:01} providing a solution concept for the geodesic (and geodesic deviation) equation for the (generalised version of the) metric~\eqref{distrib-0}, which is obtained by replacing the Dirac delta by a so-called {\bf generalized delta function}.\footnote{Technically, this is an equivalence class of smooth functions represented by a model delta net.} This setting, based on Colombeau's construction of algebras of generalized functions \cite{Col:85, Col:92}, allows for a consistent treatment of products of distributions in (semi-)Riemannian geometry. In particular, this allowed to clarify the meaning of the Penrose junction conditions and the equivalence of the distributional and continuous forms of the metric \cite{KunzingerSteinbauer:1999b}.

\begin{figure}[h]%
\centering
\includegraphics[scale=0.45]{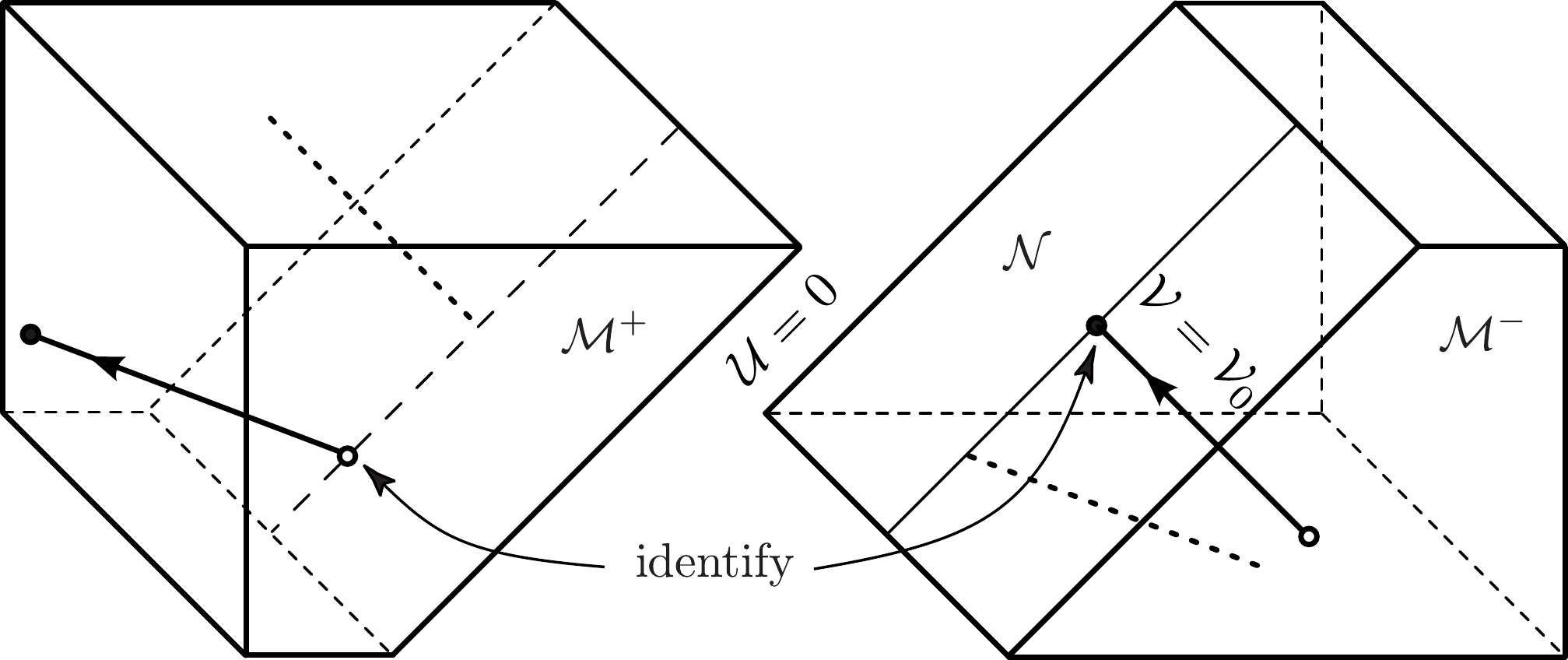}
\caption{In usual coordinates \eqref{backgr-Mink} of Minkowski space, any null geodesic with constant ${\V_0, \eta_0}$ for $\U<0$ in ${\cal M}^-$ cross the impulse ${\cal N}$ located at $\U=0$, and continues for $\U>0$ in ${\cal M}^+$. It experiences a unique ``shift'' and ``refraction'' which is exactly characterized by the Penrose junction conditions \eqref{Penrose-junction-conditions} in the distributional Brinkmann form \eqref{distrib-0} of the implusive metric. Moreover, the family of such null geodesics gives the ``comoving'' Rosen coordinates~\eqref{conti-0}.}
\label{fig3}
\end{figure}

The geometric key idea was to employ a privileged (natural) family of {\bf null geodesics} which {\bf cross the impulse}. Indeed, the existence (and uniqueness) of such geodesics was proven, enabling a study of their interaction with the impulse, see Figure~\ref{fig3}. Their ``shift'' and their ``refraction'' can explicitly be expressed in terms of the profile ${\h}$ and its derivatives at the point the corresponding geodesic hits the impulse. Finally, this family of null geodesics gives the ``comoving'' Rosen coordinates, and hence allows to smoothly and explicitly transform the generalized version of~\eqref{distrib-0} to the generalized version of~\eqref{conti-0}.

Neglecting the mathematical details, we can {\bf summarize that:}
\begin{itemize}
\item  a suitable \emph{family of null geodesics} in the distributional \emph{pp}-wave metric~\eqref{distrib-0} is explicitly constructed,
\item  they \emph{cross the impulse}, they are \emph{complete}, and \emph{define the comoving coordinates} of the Rosen metric~\eqref{conti-0}, hence
\item  they allow us to \emph{``geometrically regularize''} the discontinuous transformation~\eqref{trans}.
\end{itemize}

Put somewhat more vividly, it turns out that within nonlinear distributional geometry the impulsive \emph{pp}-wave spacetime can be equivalently described by two metrics, the generalized distributional Brinkmann form and the generalized continuous Rosen form which are related by a generalized coordinate transform. The distributional limits of the respective metrics are precisely \eqref{distrib-0} and \eqref{conti-0}, and the distributional limit of the corresponding transformation is the \emph{discontinuous transformation} \eqref{trans}, which explicitly encodes the Penrose junction conditions \eqref{Penrose-junction-conditions}, see also \cite{EG:11,Erl:13} and the diagram in~\cite[p.\ 1261]{KunzingerSteinbauer:1999b}.

Finally, we mention that this way of dealing with the intricacies of low-regularity metrics in the context of impulsive plane waves has recently lead to the clarification of a lapse in the literature on the wave memory effect in these geometries~\cite{Ste:19}.

\section{Nonexpanding impulsive gravitational waves with a cosmological constant $\Lambda$}
\label{sec4}

The \emph{pp}-waves, representing gravitational waves with plane surfaces, are solutions of Einstein's field equations \emph{only in Minkowski space}. However, their generalizations to a nonzero value of the cosmological constant~$\Lambda$ exist within the large class of {\bf Kundt spacetimes} \cite{Kundt:1961} which is defined by admitting the existence of a congruence of null geodesics without twist, shear and expansion \cite{GriffithsPodolsky:2009, SKMHH:2003}. In this context, nonexpanding impulses \emph{in de Sitter and anti-de Sitter spaces} can be constructed. In fact, they have been systematically studied since 1990s. We will now summarize the main results and the references.

\subsection{The background de Sitter and anti-de Sitter spaces}
\label{subsec4.1}

Let us begin by recalling the (anti-)de Sitter geometries. They are constant-curvature spaces which are also maximally symmetric (admitting 10 Killing vector fields), and conformally flat vacuum solutions of Einstein's equations with~$\Lambda$.  Their metric can be very conveniently written as
\begin{equation}\label{backgr}
  \dd s_0^2 = \frac{2\,\dd\eta\,\dd\bar\eta \,-\,2 \,\dd\U\,\dd\V}
     {[\,1+{\frac{1}{6}}\Lambda(\eta\bar\eta-{\U}{\V})\,]^2}\,,
\end{equation}
which nicely represents Minkowski, de Sitter, and anti-de Sitter space, for ${\Lambda=0}$, ${\Lambda>0}$, and ${\Lambda<0}$, respectively, and is explicitly conformally flat.


\begin{figure}[h]%
\centering
\includegraphics[scale=0.7]{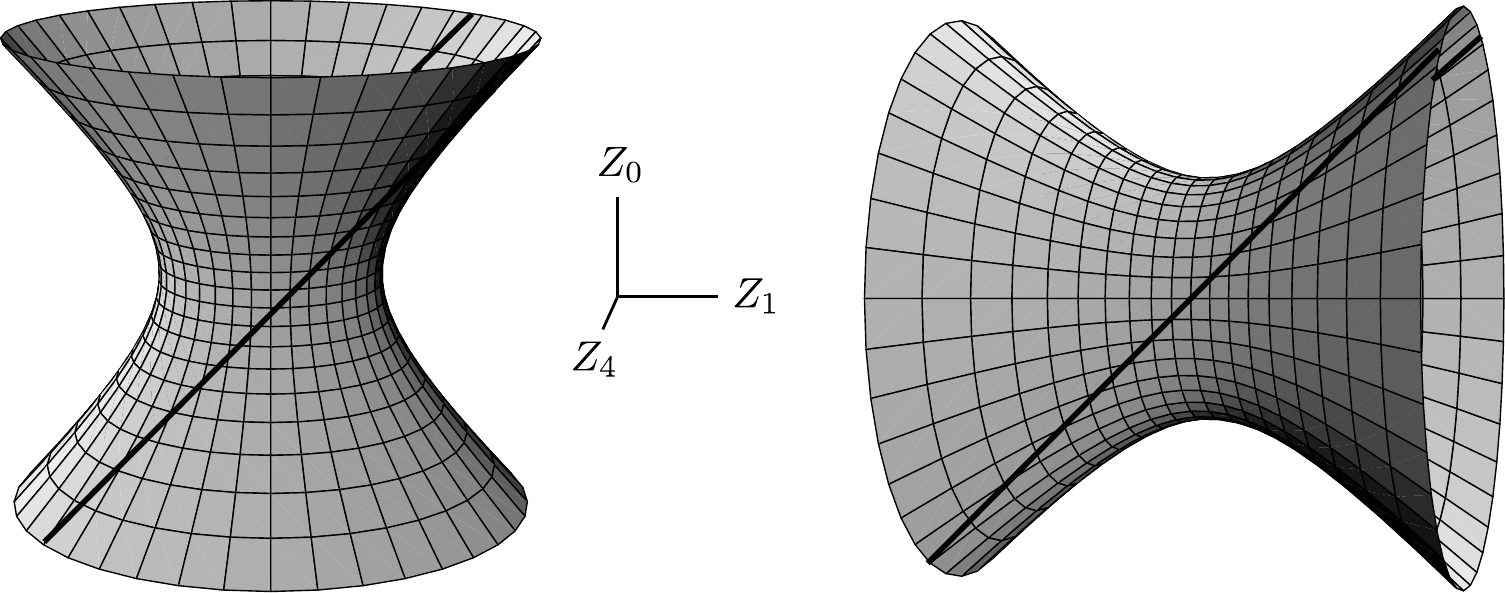}
\caption{The de Sitter (left) and anti-de Sitter (right) spaces depicted as 4-dimensional hyper\-boloids \eqref{Constraint_Hyp} embedded in 5-dimensional flat space \eqref{5D}, shown here in the section ${Z_2=0=Z_3}$. The location of the nonexpanding impulse is indicated by a pair of null lines ${Z_0=Z_1}$, that is ${U=0}$ equivalent to ${\U=0}$.}
\label{fig4}
\end{figure}

The coordinates of \eqref{backgr} cover the full {\bf (anti-)de Sitter hyperboloid}
\begin{align}
 {Z_2}^{2}+{Z_3}^{2}+ \sigma {Z_4}^{2} -2UV = \sigma\, a^{2} \label{Constraint_Hyp} \,,
\end{align}
embedded {\bf in 5-dimensional flat space}
\begin{equation}\label{5D}
  \dd s^{2} = \dd Z_{2}^{2}+\dd Z_{3}^{2}+ \sigma\dd Z_{4}^{2} -2\dd U \dd V\,,
\end{equation}
where
\begin{equation}\label{defUV}
U = \tfrac{1}{\sqrt{2}}\,(Z_0 - Z_1)\,,
\qquad
V = \tfrac{1}{\sqrt{2}}\,(Z_0 + Z_1)\,,
\end{equation}
and
\begin{equation}\label{defsigma-and-a}
\sigma \equiv \mathrm{sign}\, \Lambda =\pm 1\,,
\qquad
a=\sqrt{3/(\sigma\Lambda)}\,.
\end{equation}

The {\bf global coordinate parameterization} of the (anti-)de Sitter manifold determined by \eqref{Constraint_Hyp} in \eqref{5D} corresponding to the conformally flat metric \eqref{backgr} is
\begin{equation}
{U} = \frac{\U}{\Omega}\,, \qquad   {V} = \frac{\V}{\Omega} \,, \qquad
  Z_2+\imag\,Z_3= \sqrt2\,\frac{\eta}{\Omega}\,,
 \qquad Z_4 = \frac{2}{\Omega}\,a - a \,, \label{CoordTrans_4D_to_5D}
\end{equation}
where the conformal factor is
\begin{equation}\label{Omega}
\Omega=1+{\textstyle\frac{1}{6}}\Lambda(\eta\bar\eta-\U\V)\,.
\end{equation}
Recalling ${\sqrt2\,\eta \equiv x+ \imag\, y}$, see \eqref{Mink-null-coordinates}, we obtain ${Z_2=x/\Omega}$ and ${Z_3=y/\Omega}$. The inverse expressions to~\eqref{CoordTrans_4D_to_5D} are thus simply given by
\begin{equation} \label{trasf-inverse}
\U=\Omega\, U \,, \qquad   \V=\Omega\, V \,, \qquad x=\Omega\, Z_2 \,, \qquad y=\Omega\,
Z_3 \,,
\end{equation}
with ${\Omega=2a/(Z_4+a)}$.

\subsection{The 5-dimensional embedding formalism}
\label{subsec4.2}

There exists an interesting geometric method by which impulsive gravitational waves can be constructed in de Sitter and anti-de Sitter spaces. This was introduced in the work of Hotta and Tanaka in 1993 \cite{HottaTanaka:1993} (which also employed the ``shift function'' method by Dray and 't~Hooft \cite{DraytHooft:1985}), and further developed in \cite{PodolskyGriffiths:1997, PodolskyGriffiths:1998, HorowitzItzhaki:1999}. It employs the 5-dimensional representation \eqref{Constraint_Hyp}, \eqref{5D} of the background (anti-)de Sitter manifold.

In fact, the use of the $5$-dimensional formalism initially occurred in the works \cite{HottaTanaka:1993, PodolskyGriffiths:1997, PodolskyGriffiths:1998} where \emph{special} impulsive waves in de Sitter space were constructed by \emph{boosting the Schwarzschild--de Sitter black hole solution to the speed of light}. This itself was a natural generalization of the ultrarelativistic boost of the Schwarzschild solution, which in 1971 has lead to the famous Aichelburg--Sexl impulsive \emph{pp}-wave in Minkowski space \cite{AS:71}. In cases with any $\Lambda$, this procedure produces  specific impulsive waves generated by null particles, i.e., sources moving at the speed of light.

The embedding method is based on considering impulsive \emph{pp}-waves in flat \emph{$5$-dimensional} semi-Riemannian space~\eqref{5D}, but \emph{constraining them} to the \mbox{$4$-dimensional} hyperboloid \eqref{Constraint_Hyp}.
Specifically, we consider the {\bf \hbox{$5$-dimensional} impulsive \emph{pp}-wave}, generalizing the Brinkmann distributional metric \eqref{distrib-0},
\begin{equation}\label{5D_imp-def}
  \dd s^{2}=\dd Z_{2}^{2}+\dd Z_{3}^{2}+ \sigma\dd Z_{4}^{2} -2\,\dd U \dd V
  +H(Z_{2},Z_{3},Z_{4})\,\delta(U)\,\dd U^{2} \,.
\end{equation}
By {\bf applying the constraint \eqref{Constraint_Hyp}}, the manifold is reduced to the 4-dimen\-sional (anti-)de Sitter space whenever ${U\ne0}$. However, due to the presence of $\delta(U)$ in \eqref{5D_imp-def}, the distributional curvature arises at the single wave-surface ${U=0}$, which represents the {\bf impulse in de Sitter or anti-de Sitter space}. In view of \eqref{CoordTrans_4D_to_5D}, such an impulse is {\bf located at}
\begin{equation}\label{5D_imp-location}
\U=0\  \Leftrightarrow\  U=0\  \Leftrightarrow\ Z_0=Z_1\,.
\end{equation}
The location of the impulse is indicated in Fig.~\ref{fig4} as a pair of null lines. Moreover, from \eqref{Constraint_Hyp} we find its geometry to be given by
\begin{equation}
  {Z_2}^2+{Z_3}^2+\sigma{Z_4}^2=\sigma a^2\,.
\label{imp_surface}
\end{equation}
This is clearly a {\bf sphere $S^2$ in de Sitter space} (since ${\sigma=1}$ for ${\Lambda>0}$), and a {\bf hyperboloid $H^2$ in anti-de Sitter} space (since ${\sigma=-1}$ for ${\Lambda<0}$). Moreover, their ``radius'' is determined by the \emph{constant}~${a=\sqrt{ 3/(\sigma\Lambda)}}$, which means that such an impulse is indeed {\bf nonexpanding}.

\subsection{Continuous, distributional, and embedding metric forms of impulses in (anti-)de Sitter space}

The impulsive waves discussed above, be they constructed by embedding the 4-dimensional (anti-)de Sitter hyperboloid \eqref{Constraint_Hyp} into the  $5$-dimensional \emph{pp}-wave spacetime \eqref{5D_imp-def}, or by considering an ultrarelativistic limit of special sources, {\bf belong to the Kundt class of spacetimes}.

All Kundt-type solutions of vacuum Einstein's equations with ${\Lambda\ne0}$ of algebraic type~N (i.e., those which represent ``pure'' gravitational waves) were explicitly found in \cite{GarciaPlebanski:1981, OzsvathRobinsonRozga:1985}. In these works it was demonstrated that there are \emph{two distinct subclasses} of such spacetimes when ${\Lambda=0}$,  \emph{only one subclass} when ${\Lambda>0}$, but (interestingly) \emph{three subclasses} when ${\Lambda<0}$, including a special Siklos solution \cite{Siklos:1985}, representing gravitational waves with hyperbolic surfaces in the anti-de~Sitter universe (see \cite{Podolsky:1993, Podolsky:1998Siklos, BicakPodolsky:1999a}, or the review \cite{GriffithsPodolsky:2009} for more details concerning the classification and mutual relations between the subclasses).

These {\bf various subclasses} are geometrically distinct for \emph{a general (smooth) wave-profile}. However, in the {\bf impulsive case}, i.e., if the wave profile is taken to be a Dirac delta thus localizing the Kundt waves to \emph{just one}  impulsive wave surface, only a {\bf single (locally) unique} impulsive class of solutions exists --- a surprising result proved by the first author in \cite{Podolsky:1998}.

A number of questions thus naturally arose, namely: Can such nonexpanding impulsive waves with $\Lambda$, which possess a unique \emph{distributional} form, also be written in some \emph{continuous} metric form?
What are their mutual \emph{coordinate relations}?
Can the Penrose \emph{``cut-and-paste''} method be extended to such a cosmological setting?
And, what are the corresponding \emph{Penrose junction conditions generalized to any cosmological constant~$\Lambda$}?

These questions were basically answered in 1999 in the work \cite{PodolskyGriffiths:1999} of the first author with Jerry Griffiths.
The trick was to employ the convenient metric~\eqref{backgr} which describes the full (anti-)de Sitter background space via the simple parametrization \eqref{trasf-inverse}, \eqref{Omega} of the hyperboloid \eqref{Constraint_Hyp}. Moreover, the metric~\eqref{backgr} is {\bf conformally flat}, so the only difference with respect to impulsive \emph{pp}-waves is the presence of the overall {\bf conformal factor} ${\Omega=1+\tfrac{1}{6}\,\Lambda(\eta\bar\eta-\U\V)}$. This approach allowed us to generalize in \cite{PodolskyGriffiths:1999} the continuous Rosen form \eqref{conti-0} and the distributional Brinkmann form \eqref{distrib-0} of impulsive \emph{pp}-waves, to find their mutual transformation, and to obtain the relation to the embedding metric \eqref{5D_imp-def}. The results are as follows:
\vspace*{2mm}

Nonexpanding impulsive gravitational waves in (anti-)de Sitter space can be written in the following {\bf 3 alternative forms:}
\begin{itemize}
\item  A {\bf continuous} metric
\begin{equation}
\mbox{\fbox{
{
$\displaystyle \dd s^2 = \frac{2\,\big\lvert\dd Z+\up(u)(h_{,\bar Z Z}\,\dd Z+{h}_{,\bar Z\bar Z}\,\dd\bar Z)\big\lvert^2
  -2\,\dd u\,\dd v}{\big[\,1+\frac{1}{6}\Lambda(Z\bar Z-uv+\up(u)\,G)\,\big]^2}\,,$
}
}}\label{conti}
\end{equation}
where $\up(u)$ is the \emph{kink function}~\eqref{kink-Heaviside}, ${h\equiv h(Z, \bar Z)}$ is any real-valued function as in~\eqref{h}, and the function $G(Z,\bar Z)$ is defined as ${G\equiv Z\,h_{,Z}+\bar Z\,h_{,\bar Z}-h}$. Again, the non-differentiability of $\up$ gives rise to a $\delta$-function curvature located at $u=0$.
For ${\Lambda=0}$ the metric~\eqref{conti} clearly reduces to the \emph{Rosen form} of impulsive \emph{pp}-waves  \eqref{conti-0}.
\vspace*{2mm}

\item A {\bf distributional} metric
\begin{equation}
\mbox
{\fbox{
{
$\displaystyle \dd s^2 = \frac{2\,\dd\eta\,\dd\bar\eta-2\,\dd \U\,\dd \V+2\,\h(\eta,\bar\eta)\,\delta({\U})\,\dd {\U}^2}
{[\,1+\frac{1}{6}\Lambda(\eta\bar\eta-{\U}{\V})\,]^2}\,,$
}
}}\label{distrib}
\end{equation}
where $\delta(\U)$ is the \emph{Dirac delta distribution}. The corresponding impulse in (anti-)de Sitter space \eqref{backgr} is thus clearly located at ${\U=0}$, and its geometry is
${\,2\,\dd\eta\,\dd\bar\eta\,[\,1+\frac{1}{6}\Lambda\,\eta\bar\eta\,]^{-2}}$, that is a sphere $S^2$ in de Sitter space and a hyperboloid $H^2$ in anti-de Sitter space, in full agreement with \eqref{imp_surface}. For ${\Lambda=0}$ we immediately recover planar impulsive wave in Minkowski space, namely the \emph{Brinkmann form} of impulsive \emph{pp}-waves \eqref{distrib-0}.
\vspace*{2mm}

\item  An {\bf embedding} metric, as introduced in Section~\ref{subsec4.2}, namely
\vspace{-4mm}

\begin{equation}
\mbox
{\fbox{
{
$\dd s^{2} = \dd Z_{2}^{2}+\dd Z_{3}^{2}+\sigma\dd Z_{4}^{2}-2\dd U \dd V
  +H(Z_{2},Z_{3},Z_{4})\,\delta(U)\,\dd U^{2}\,, $
}
}}\label{5D_imp}
\end{equation}
with the {\bf constraint} \eqref{Constraint_Hyp}, that is
\begin{equation}
\mbox
{\fbox{
{
${Z_2}^{2}+{Z_3}^{2}+ \sigma {Z_4}^{2} -2UV = 3/\Lambda\,, $
}
}}\label{Constraint_Hyp-again}
\end{equation}
where ${\sigma = 1}$ for ${\Lambda>0}$ while ${\sigma = -1}$ for ${\Lambda<0}$.
This is the 5-dimensional \emph{pp}-wave metric in the distributional (Brinkmann) form constrained to the
(anti-)\-de Sitter hyperboloid \eqref{Constraint_Hyp-again}. The nonexpanding impulse is located at ${U=0 \Leftrightarrow Z_0=Z_1}$, see \eqref{5D_imp-location} and Fig.~\ref{fig4}, and its geometry is given by \eqref{imp_surface}.

\end{itemize}

\subsection{Relating the alternative metric forms}

It is now straightforward to formally obtain the coordinate relation \emph{between the 5-dimensional embedding metric} \eqref{5D_imp} \emph{and the distributional metric} \eqref{distrib}. By applying the {\bf parametrization} \eqref{CoordTrans_4D_to_5D} of the (anti-)de Sitter hyperboloid  --- which satisfies the constraint \eqref{Constraint_Hyp-again} --- to the metric \eqref{5D_imp}, the  (anti-)de Sitter background part $\dd s_0^{2}$ takes the form \eqref{backgr}, and there is an additional impulsive term ${H\,\delta(U)\,\dd U^{2}}$. Because ${{U} = \U\,\Omega^{-1}}$, using the distributional identities ${\delta(U)=\Omega\,\delta(\U)}$ and ${\U\delta(\U)=0}$ it can be written as ${H\,\Omega^{-1}\,\delta({\U})\,\dd {\U}^2}$. This is exactly the last term ${2\,\h\,\Omega^{-2}\,\delta({\U})\,\dd {\U}^2}$ in the distributional metric \eqref{distrib}, with the obvious identification
\begin{equation}\label{H-relations}
  \h = \tfrac{1}{2}\,H\,(1+\tfrac{1}{6}\Lambda\,\eta\bar\eta)\,.
\end{equation}

Similarly, we can find the transformation \emph{between the distributional metric} \eqref{distrib} \emph{and the continuous metric} \eqref{conti}. Since their numerators are exactly the Brinkmann and Rosen forms of impulsive \emph{pp}-waves, respectively, which are related by the transformation \eqref{trans}, we only need to compare their conformal factors. Applying the relations \eqref{trans} and the identity (of $L^\infty$-functions) ${\up=u\,\Theta}$ we observe that actually
\begin{equation}
\Omega = 1+\tfrac{1}{6}\Lambda\big(\eta\bar\eta-\U\V\big)
       = 1+\tfrac{1}{6}\Lambda\big(Z\bar Z-uv+\up(u)\, G\big)\,.
\end{equation}
We thus conclude that the metric forms \eqref{distrib} and \eqref{conti} of nonexpanding impulses in (anti-)de Sitter~space are {\bf related by the same  discontinuous transformation} \eqref{trans} {\bf as in flat space}, namely

\begin{equation}
\fbox{\begin{minipage}{5.6cm}
\vspace{-0.5cm}
  \begin{eqnarray}
  \  {\U} &=& u \,, \nonumber \\
  \  {\V} &=& v+ \Theta(u)\,h+\up(u)\,h_{,Z}\,h_{,\bar Z} \,,\nonumber\\
  \  \eta &=& Z+\up(u)\,h_{,\bar Z} \,.\nonumber
  \end{eqnarray}
\vspace{-0.7cm}
\end{minipage}}
\label{trans-again}
\end{equation}
\vspace{1.5mm}

This fact enables us to generalize the Penrose ``cut-and-paste'' construction method and to formulate the corresponding Penrose junction conditions for any value of the cosmological constant~$\Lambda$.

\subsection{Penrose junction conditions with $\Lambda$}

The transformation between the continuous metric \eqref{conti} and the distributional metric \eqref{distrib} is given by the discontinuous transformation \eqref{trans-again}, which is independent of $\Lambda$. The Penrose {\bf ``cut-and-paste'' method} for construction of these nonexpanding impulsive waves can thus be used similarly as in flat Minkowski space ${\cal M}$, provided the more general, conformally flat metric \eqref{backgr} is employed. The only difference is that the {\bf cut is performed in de Sitter manifold}~$d{\cal S}$ or {\bf anti-de Sitter manifold~$Ad{\cal S}$} along the null surface ${\cal N}$, expressed as ${\,\U=0}$ in the background coordinates of \eqref{backgr}. Such a cut is indicated in Fig.~\ref{fig4}. The {\bf pasting} is then performed using the {\bf Penrose junction conditions} at ${\U=0}$
\begin{equation}
\mbox{\fbox{
{$\big(\V,\eta,\bar\eta\big)_{{(A)d{\cal S}}^-}\,=\,\big(\V-\h(\eta,\bar\eta),\eta,\bar\eta\big)_{{(A)d{\cal S}}^+}\,, $}
}}
\label{Penrose-junction-conditions-again}
\end{equation}
which are {\bf formally the same as} \eqref{Penrose-junction-conditions} in the original case ${\Lambda=0}$. The geometric reason is that the cosmological constant enters the metric \eqref{backgr} only via the conformal factor.

Moreover, the Penrose junction conditions \eqref{Penrose-junction-conditions-again} are again {\bf implicitly encoded in the transformation} \eqref{trans-again}, namely via the deformed shift along $\V$ given by the real function $\h(\eta,\bar\eta)$ which corresponds to the step-term $\Theta(u)\,h$ therein. Indeed, it follows from \eqref{trans-again} that ${\h(\eta,\bar\eta)=h(Z, \bar Z)}$ because ${\eta = Z}$ at ${u=\U=0}$ due to the continuity of the kink function $\up(u)$.

\section{Relating the continuous and distributional metric forms with $\Lambda$}
\label{sec5}

Above we have argued that the continuous metric \eqref{conti} and the distributional metrics \eqref{distrib} and \eqref{5D_imp} with \eqref{Constraint_Hyp-again} are equivalent. But from the mathematical point of view these relations are only \emph{formal}, again due to the discontinuity of the transformation \eqref{trans-again}. This is analogous to the situation in impulsive \emph{pp}-waves concerning the relation between the continuous Rosen form \eqref{conti-0} and the distributional Brinkmann form \eqref{distrib-0}. As in Section~\ref{subsec3.2}, to find the exact relation between the metrics, and to rigorously prove their equivalence, \emph{it is necessary to study the geodesics}.

\subsection{Geodesics in nonexpanding impulsive waves with $\Lambda$}
\label{subsec5.1}

All geodesics crossing any nonexpanding impulse in (anti-)de Sitter space were investigated and (formally) found in 2001 by the first author and Marcello Ortaggio \cite{PodolskyOrtaggio:2001}. Using the embedding formalism, they are explicitly given by
\begin{equation}\label{5DNullGeod2001U}
 U=t\,,\qquad U=a\dot U^0\sinh(t/a)\,,\qquad U=a\dot U^0\sin(t/a)\,,
\end{equation}
in the cases ${\sigma e=0}$, ${\sigma e<0}$, and  ${\sigma e>0}$, respectively. Here ${\sigma = \mathrm{sign}\, \Lambda}$, and $e$ determines the velocity normalization (${e=-1}$ for timelike, ${e=0}$ for null, and ${e=1}$ for spacelike geodesics, respectively). The affine parameter $t$ is chosen in such a way that each geodesic crosses the impulse ${U=0}$ at ${t=0}$.

Without loss of generality, $\dot U^0$ can be taken to be positive, so  that \eqref{5DNullGeod2001U} are increasing functions. \emph{Using $U$ as a convenient parameter} and applying some distributional identities, the {\bf general solution} of the remaining geodesic equations can be derived. For {\bf null geodesics} with ${\dot U^0=1}$, it can be written\footnote{See Sec.~5 in \cite{SaemannSteinbauerLeckePodolsky:2016}, namely Proposition~5.3. Relation to the original form presented in \cite{PodolskyOrtaggio:2001} (which uses a different definition of~$C$) is contained in Remark~5.4.}
\begin{align}\label{5DNullGeod2001}
Z_p(U)&= Z_p^0 + \dot{Z}^0_p\,U + A_p\,U\,\Theta(U) \,, \nonumber\\
V(U)  &= V^0   + \dot{V}^0\,U   + B\,\Theta(U) + C\,U\,\Theta(U)\,.
\end{align}
The constants ${Z_p^0, \dot{Z}^0_p, V^0, \dot{V}^0}$ for ${p=2,3,4}$ are determined by the initial data, while the coefficients $A_p, B, C$ are
 \begin{align}\label{eq:ro:final}
 A_j &= \tfrac{1}{2}\Big(H_{,j}(0)-\frac{\Lambda}{3}\,Z^0_j\,G(0)\Big)  \qquad \hbox{for}\ j=2,3\,, \nonumber\\
 A_4 &= \tfrac{1}{2}\Big(\sigma H_{,4}(0)-\frac{\Lambda}{3}\,Z^0_4\,G(0)\Big)\,, \qquad
 B   = \tfrac{1}{2}\,H(0), \nonumber\\[1mm]
 C   &= \tfrac{1}{8} \Big( H_{,2}(0)^2 + H_{,3}(0)^2 + \sigma H_{,4}(0)^2 
 + \frac{\Lambda}{3}\,H(0)^2 - \frac{\Lambda}{3}\left( \delta^{pq}\,Z_p^0\, H_{,q}(0)\right)^2 \Big)  \nonumber\\
     &\quad -\frac{\Lambda}{6}\,V^0\,G(0)
     +\tfrac{1}{2}\delta^{pq}\,\dot{Z}^0_p\, H_{,q}(0)\,,
 \end{align}
${H(0)= H(Z^0_p)}$,  ${G(0)=G(Z_p^0)\equiv\delta^{pq}\,Z_p^0\, H_{,q}(Z_r^0)-H(Z_r^0)}$, ${H_{,q}(0)= H_{,q}(Z^0_p)}$ are functions evaluated on the impulse ${U=0}$, given by $H(Z_{2},Z_{3},Z_{4})$ of \eqref{5D_imp}.

Hence we see that, analogous to the case of impulsive \emph{pp}-waves, the global geodesics suffer a \emph{jump in} $V$ across the impulse (determined by $B$) and, in addition, are \emph{broken} in the $V$ and $Z_p$-directions (determined by $C$ and $A_p$, respectively). The coefficients specifying the magnitude of these effects are again given by the profile function $H$ and its derivatives, evaluated at the points where the respective geodesic hits the impulse.
\vspace*{2mm}

These results were {\bf fully confirmed} in 2016 \cite{SaemannSteinbauerLeckePodolsky:2016, SaemannSteinbauer:2017} by a rigorous investigation of the geodesic equations using a {\bf regularisation technique}. Again the Dirac $\delta$ in the metric, now in \eqref{5D_imp}, was replaced by a model delta nets $\delta_\varepsilon$. In physical terms this  means that the formal distributional form of the impulsive metric is understood as a limit of a family of spacetimes with ever shorter but stronger sandwich gravitational waves with smooth profile $\delta_\varepsilon$.

Although the resulting geodesic equation for the regularized metric(s) forms a highly coupled system, it was possible to prove the existence and uniqueness of geodesics that (at least for small regularization parameters $\varepsilon$) exist long enough such that they {\bf cross the regularized wave impulse}, i.e., the region $-\varepsilon\leq U\leq \varepsilon$. Since off that region we are dealing with the (anti-)de Sitter background this immediately {\bf implies completeness} of the geodesics of the regularized spacetime(s).
The proof is based on an \emph{application of Weissinger's fixed point theorem} \cite{Wei:52}. Remarkably, when taking the impulsive limit ${\varepsilon\to0}$, the geodesics converge to the \emph{the same limit for any choice of profile $\rho$} of the sandwich gravitational waves, i.e., they are independent of the specific regularization. Moreover, they fully agree with \eqref{5DNullGeod2001}--\eqref{eq:ro:final}, and extend these formulae to \emph{any} value of the initial data ${\dot{U}^0, Z_p^0, \dot{Z}^0_p, V^0, \dot{V}^0}$.
\vspace*{2mm}

Moreover, we have also investigated the geodesics in these spacetimes {\bf employing the continuous form of the metric}\,~\eqref{conti}. This analysis is based on the local Lipschitz continuity of this metric and the observation made in \cite{Steinbauer:2014} that {\bf every locally Lipschitz metric has $C^1$-geodesics} in the sense of Filipov \cite{Fil:88}. This solution concept for ordinary differential equations with discontinuous right-hand sides was also used in~\cite{PodolskySaemannSteinbauerSvarc:2015} to establish the existence and uniqueness of continuously differentiable geodesics crossing the wave impulse. We also explicitly derived their form using a $C^1$-matching procedure.
\vspace*{2mm}

A natural question thus arose about the {\bf mutual consistency} of these two results,
both obtained in a rigorous way but starting from two different forms of the
metric, namely the continuous
and the embedding $5$-dimensional distributional form.
It can indeed \emph{formally} be shown that the form of the geodesics derived in both ways are the same
when appropriate coordinate transformations are applied. This result confirmed that both our approaches are consistent, and that the understanding of geodesics in the complete family of spacetimes with
nonexpanding impulsive gravitational waves and any cosmological constant now
rests on firm mathematical grounds, preparing us for the next step.

\subsection{Deriving the discontinuous transformation}
\label{subsec5.2}

With these explicit geodesics at hand, we were able in \cite{PodolskySaemanSteinbauerSvarc:2019} to geometrically {\bf derive} the discontinuous transformation \eqref{trans-again} and the Penrose junction conditions \eqref{Penrose-junction-conditions-again}, thus clarifying their nature and meaning, and laying the foundations for their rigorous treatment analogous to the one for the \emph{pp}-wave case in \cite{KunzingerSteinbauer:1999b}. Let us summarize the key steps and the main results:
\vspace*{2mm}

\begin{itemize}

\item  First, in the distributional metric \eqref{distrib} with coordinates ${(\U, \V, x, y)}$ we identify a {\bf special family of global null geodesics}  which cross the impulse located at ${\U=0}$. We get them by {\bf naturally setting their initial values} ${\V_0, x_0, y_0}$ to be constants (with no velocity) in (anti-)de Sitter space ($A$)${{d{\cal S}}^-}$ {\bf in front of the impulse}, that is for ${\U<0}$. So in front of the impulse they are the generators of the (anti-)de Sitter space, when expressed in the $5$-dimensional flat space.

\vspace*{2mm}
\item  Such {\bf null geodesics are} explicitly given by
\begin{equation}
\gamma({\lambda}) = \Big( \frac{\alpha\lambda}{1-\beta\lambda}, \V_0, x_0, y_0 \Big)\,,
\end{equation}
where $\lambda$ is the affine parameter and
\begin{equation} \label{4DGeodAfBetaNeq0}
\alpha\equiv 1+\frac{\Lambda}{12}\left(x_0^2+y_0^2\right) \,, \qquad
\beta\equiv -\frac{\Lambda}{6}\,\V_0\,.
\end{equation}
Notice that ${\U=0}$ iff ${\lambda=0}$.

\vspace*{2mm}
\item  Next we express these null geodesics in the {\bf $5$-dimensional embedding} form of (anti-)de Sitter space in the coordinates ${(U,V,Z_2,Z_3,Z_4)}$ of \eqref{5D} using \eqref{CoordTrans_4D_to_5D} as
\begin{equation}\label{5DGeodFin}
\gamma(\lambda) = \frac{1-\beta\lambda}{\alpha}\,
\bigg(
  \frac{\alpha\lambda}{1-\beta\lambda}, \V_0, x_0, y_0, a\Big(2-\frac{\alpha}{1-\beta\lambda}\Big)
\bigg)\,.
\end{equation}
Notice that ${\,U=\lambda}$.

\vspace*{2mm}
\item  Now we employ the results on {\bf global null geodesics} crossing the impulse at ${U=0}$ to the region ${U>0}$ behind it, expressed {\bf in the $5$-dimensional coordinates} $(U,V,Z_2,Z_3,Z_4)$. As explained in Section \ref{subsec5.1}, these were rigorously derived using a {\bf general regularisation}. After a distributional limit they take the form

\begin{equation}\label{5DNullGeod}
\gamma_{5D}(\lambda) \equiv \left(
\begin{array}{c}
U(\lambda)\\
V(\lambda)\\
Z_p(\lambda)
\end{array}\right) =\left(
\begin{array}{c}
\lambda\\
  V^0+\dot{V}^0\,\lambda+B\,\Theta(\lambda)+C\,\lp(\lambda)\\
Z_p^0+\dot{Z}^0_p\,\lambda+A_p\,\lp(\lambda)
\end{array}\right),
\end{equation}
\vspace*{1mm}

\noindent
cf.\ \eqref{5DNullGeod2001}, \eqref{eq:ro:final}, with the {\bf initial data} (for ${U<0}$)
\begin{align}\label{GenPos}
&V^0=\frac{\V_0}{\alpha} \,,& &Z_2^0=\frac{x_0}{\alpha} \,,&
&Z_3^0=\frac{y_0}{\alpha} \,,& &Z_4^0=a\,\Big(\frac{2}{\alpha}-1\Big) \,,
\nonumber \\
&\dot{V}^0=-\frac{\beta}{\alpha}\,\V_0 \,,&
&\dot{Z}^0_2=-\frac{\beta}{\alpha}\,x_0 \,,& &\dot{Z}^0_3=-\frac{\beta}{\alpha}\,y_0
\,,& &\dot{Z}^0_4=-2a\,\frac{\beta}{\alpha} \,. 
\end{align}
Here $\lambda_+(\lambda)\equiv\lambda\,\Theta(\lambda)$, while the coefficients are
\begin{align}
& A_j=\h^\imag_{,j}+\frac{x_0^j}{2\sigma\alpha a^2}\, \G, \quad 
  A_4=\frac{1}{\sigma\alpha a} \G, \quad 
  B = \frac{1}{\alpha}\h^\imag \,,\nonumber \\
& C =\frac{1}{2}\big((\h^\imag_{,x})^2+(\h^\imag_{,y})^2\big)+\frac{1}{
2\sigma\alpha a^2}\big((\h^\imag+\G)\V_0+\h^\imag\G\big) \,, \label{5DCoeffSpec}
\end{align}
with ${\G \equiv \h^\imag-x_0\h^\imag_{,x}-y_0\h^\imag_{,y}}$. Here the superscript  $\,^\imag$ denotes the evaluation of the respective functions at ${U=0}$, i.e., ``at the impulse''.

\vspace*{2mm}
\item Next we express these null geodesics {\bf in the distributional coordinates}
$(\U,\V,x,y)$ for the (anti-)de Sitter space ${{(A)d{\cal S}}^+}$ {\bf behind the impulse}, that is for ${\U>0}$, using the  relations \eqref{trasf-inverse},
\begin{equation} \label{trasf-inverse2}
\U=\Omega\, U \,, \quad   \V=\Omega\, V \,, \quad x=\Omega\, Z_2 \,, \quad y=\Omega\,
Z_3 \,, \quad \hbox{with} \ \,\Omega=\frac{2a}{Z_4+a}\,.
\end{equation}
A somewhat lengthy calculation gives the {\bf explicit global null geodesics} crossing the impulse in the form
\begin{align}
&\gamma_{4D}[\V_0,x_0,y_0](\lambda) \label{Global4DNullGeod}\\
&\ \equiv \left(\!\!
\begin{array}{c}
\U(\lambda)\\
\V(\lambda)\\
x^j(\lambda)
\end{array}\!\!\right) = \left(
\begin{array}{c}
{\displaystyle \frac{\alpha\lambda}{1-\beta\lambda+\frac{\Lambda}{6}\G\,\lp}}
\vspace{2.0mm} \\ \!\!
\V_0+{\displaystyle \frac{\Theta(\lambda)}{1-\beta\lambda+\frac{\Lambda}{6}\G\,\lp}}\,
\h^\imag+{\displaystyle \frac
{\alpha\lp}{2(1-\beta\lambda+\frac{\Lambda}{6}\G\,\lp)}}\,{\mathcal F}\!
\vspace{2.0mm}  \\
x_0^j+{\displaystyle \frac{\alpha\lp}{1-\beta\lambda+\frac{\Lambda}{6}\G\,\lp}}\,\h_{,j}^\imag\\
\end{array}\right). \nonumber
\end{align}
We have denoted the dependence on the initial data $[\V_0,x_0,y_0]$ explicitly, and we have used the abbreviation
\begin{equation}
  {\mathcal F}\equiv(\h_{,x}^\imag)^2+(\h_
  {,y}^\imag)^2+\frac{\h^\imag}{\sigma\alpha a^2}(\V_0+\G)\,.
\end{equation}
Recall also that the constants $\alpha$, $\beta$ are given by \eqref{4DGeodAfBetaNeq0}.

\vspace*{2mm}
\item Finally, using the relations
 ${\U(\lambda)=\alpha\lambda/(1-\beta\lambda+\frac{\Lambda}{6}\G\,\lp)}$ and ${\Theta(\lambda)=\Theta(\U)}$ implying ${\,\Up = \alpha\lp/(1-\beta\lambda+\frac{\Lambda}{6}\G\,\lp)}$, the result \eqref{Global4DNullGeod} can be simply rewritten as

\vspace*{2mm}

\hspace*{-4mm}
\fbox{\begin{minipage}{6.4cm}
\vspace{-3mm}

\begin{align}
\gamma_{4D}[\V_0,x_0,y_0](\U)\equiv \left(\!\!
\begin{array}{c}
\U\\
\V(\U)\\
x^j(\U)
\end{array}\!\!\right) \!=\!\left(\!\!\!
\begin{array}{c}
\U \vspace{2.0mm} \\
\V_0+\Theta(\U)\,\h^\imag+\Up
\,\frac{1}{2}\big[(\h_{,x}^\imag)^2+(\h_{,y}^\imag)^2\big] \vspace{2.0mm}  \\
x_0^j+\Up\,\h_{,j}^\imag\\
\end{array}\!\!\!\right)\!. \nonumber
\end{align}
\vspace{-2mm}
\end{minipage}}

\end{itemize}

\noindent
These is the rigorously derived explicit form of the global null geodesics of the distributional metric \eqref{distrib}, influenced by the nonexpanding impulse in (anti-)de Sitter space \eqref{backgr}.

\subsection{The discontinous transformation and the Penrose junction conditions with $\Lambda$}
\label{subsec5.3}

The explicit form of global null geodesics in (anti-)de Sitter space with nonexpanding impulsive gravitational waves just obtained {\bf can be employed as a transformation from the continuous to the corresponding distributional} form of the metric ${(u,v,X,Y) \mapsto (\U,\V,x,y)}$ with ${u=\U}$. The trick is:
\vspace*{0mm}

\hspace*{-4mm}
\fbox{\begin{minipage}{11.2cm}
\vspace*{1mm}

Take the {\bf initial values} ${\V_0,x_0,y_0}$ as the {\bf comoving coordinates}.

They define the coordinates of the continuous metric form.

\vspace*{1mm}
\end{minipage}}
\vspace*{3mm}

\noindent
In a fully explicit form we hence obtain the mapping

\vspace*{3mm}

\hspace*{-4mm}
\fbox{\begin{minipage}{11.2cm}
\vspace{-1mm}
\begin{equation}\nonumber
  \left(\!\!\!\begin{array}{c}
    u\\v\\X\\Y
   \end{array}\!\!\!\right)\!\!
   \mapsto \gamma_{4D}[v,X,Y](u)\!=\!\!
   \left(\begin{array}{c}
     u\\\!\!\!v+\Theta(u)\,\h^\imag+\up
     \,\frac{1}{2}\big[(\h_{,X}^\imag)^2+(\h_{,Y}^\imag)^2\big]\!\!\! \\
     X+\up\h_{,X}^\imag\\ Y+\up\h_{,Y}^\imag
   \end{array}\right)
   \!\!=\!\!
   \left(\!\!\begin{array}{c}
     \U\\\V\\x\\y
   \end{array}\!\!\right).
\end{equation}
\vspace{-2mm}
\end{minipage}}

\vspace*{3mm}

\noindent
This is the relation between the \emph{continuous} metric \eqref{conti} and \emph{distributional} metric \eqref{distrib}. Indeed, by denoting
\begin{equation}
Z\equiv\tfrac{1}{\sqrt{2}}(X+\imag\,Y)\,,\qquad
\eta=\tfrac{1}{\sqrt{2}}\,(x+\imag\,y)\,,\qquad
h(Z,\bar Z)\equiv \h^\imag=\h(0)\,,
\end{equation}
we get
\begin{eqnarray}\label{discontinuous-transformation}
  \quad {\U}&=&u\,,  \nonumber \\
  \quad {\V}&=&v+\Theta(u)\,h+\up(u)\,h_{,Z}\,h_{,\bar Z}\,,\\
  \quad \eta&=&Z+\up(u)\,h_{,\bar Z}\,.\nonumber
\end{eqnarray}
which is {\bf exactly the discontinuous transformation} \eqref{trans-again}.

\newpage

Moreover, by inspecting \eqref{discontinuous-transformation} it is seen that the coordinates $\eta, \bar\eta$ are continous across the impulse at ${\U=0}$, but there is a jump in $\V$ having the value $h(Z, \bar Z)=\h(\eta,\bar\eta)$, i.e., ${\V \mapsto \V - \h}$ across the impulse ${\cal N}$.  {\bf This provides
the justification --- and, in fact, a systematic derivation --- of the Penrose junction conditions \eqref{Penrose-junction-conditions-again} in de Sitter and anti-de Sitter space.}

\vspace*{3mm}

In a nutshell, our overall procedure employed in this section was:

\vspace*{2mm}

\begin{itemize}

\item We have \emph{derived} the discontinuous transformation \eqref{discontinuous-transformation} from a special family of \emph{global null geodesics} in (anti-)de Sitter space with impulsive waves in the distributional form, obtained by a \emph{general regularisation}. To do so, we employed the $5$-dimensional embedding formalism where these geodesics are easily seen to be the generators of the (anti-)de Sitter hyperboloid.

\vspace*{1mm}
\item This transformation \emph{turns these special null geodesics into coordinate lines}, hence the coordinates $(u,v,X,Y)$, equivalent to $(u,v,Z,\bar Z)$, are \emph{comoving} with the corresponding null particles.

\vspace*{1mm}
\item In fact, that is the way how the \emph{distributional metric}  \eqref{distrib} is transformed to the much more regular \emph{(locally Lipschitz) continuous metric} \eqref{conti}.

\vspace*{1mm}
\item This \emph{generalises (and is in perfect agreement with) the case} ${\Lambda=0}$,
 in which the continous Rosen coordinates \eqref{conti-0} for \emph{pp}-waves are also comoving, and are related to the distributional Brinkmann coordinates \eqref{distrib-0} by the same transformation \eqref{trans}.

\vspace*{1mm}
\item Moreover, the rigorously derived discontinuous transformation \eqref{discontinuous-transformation} \emph{justifies and proves the uniqueness of the Penrose junction conditions} \eqref{Penrose-junction-conditions-again} for construction of nonexpanding impulsive waves in de Sitter and anti-de Sitter spaces.

\vspace*{1mm}
\item Finally, this approach also allows us to extend the mathematically precise analysis of the Penrose ``discontinuous coordinate transformation'' described in Section \ref{subsec3.2} to the case of non-vanishing cosmological constant $\Lambda$. Its detailed implementation will appear in a forthcoming, more technical paper.
\end{itemize}

\section{``Cut-and-paste'' with $\Lambda$}
\label{sec6}

In this concluding section we will employ the above results to derive a nice picture generalizing the Penrose ``cut-and-paste'' picture of Figures \ref{fig1} and \ref{fig3} to any non-vanishing $\Lambda$.
Indeed, the transformation \eqref{discontinuous-transformation} demonstrates the geometrical importance of the special family of null geodesics employed, namely {\bf the null geodesic generators of the (anti-)de Sitter hyperboloid} \eqref{Constraint_Hyp-again}, which are transverse to the generator at which the wave impulse is located. For a fixed value of $\eta$, these are given simply by ${\U=\hbox{const.}}$ and ${\V=\hbox{const.}}$ in the conformally flat form of the metric \eqref{backgr} given by the parametrization \eqref{trasf-inverse}.

Specifically, due to the interaction with the impulse ${\cal N}$ located at ${\U=0}$, encoded in the transformation \eqref{discontinuous-transformation}, the null generators ${\V=\V_0=\hbox{const.}}$ suffer {\bf the jump ${\V_0 \mapsto \V_0 - \h^\imag}$ prescribed by the function $\h$}, according to the Penrose junction conditions \eqref{Penrose-junction-conditions-again}, {\bf but also a unique refraction}. The reason for it is that the corresponding fixed value of $\V$ of the \emph{global and unique} null geodesic {\bf behind the impulse must again be a null generator of the hyperboloid}.

\begin{center}
\begin{figure}[h]%
\hspace{15mm}\includegraphics[scale=0.48]{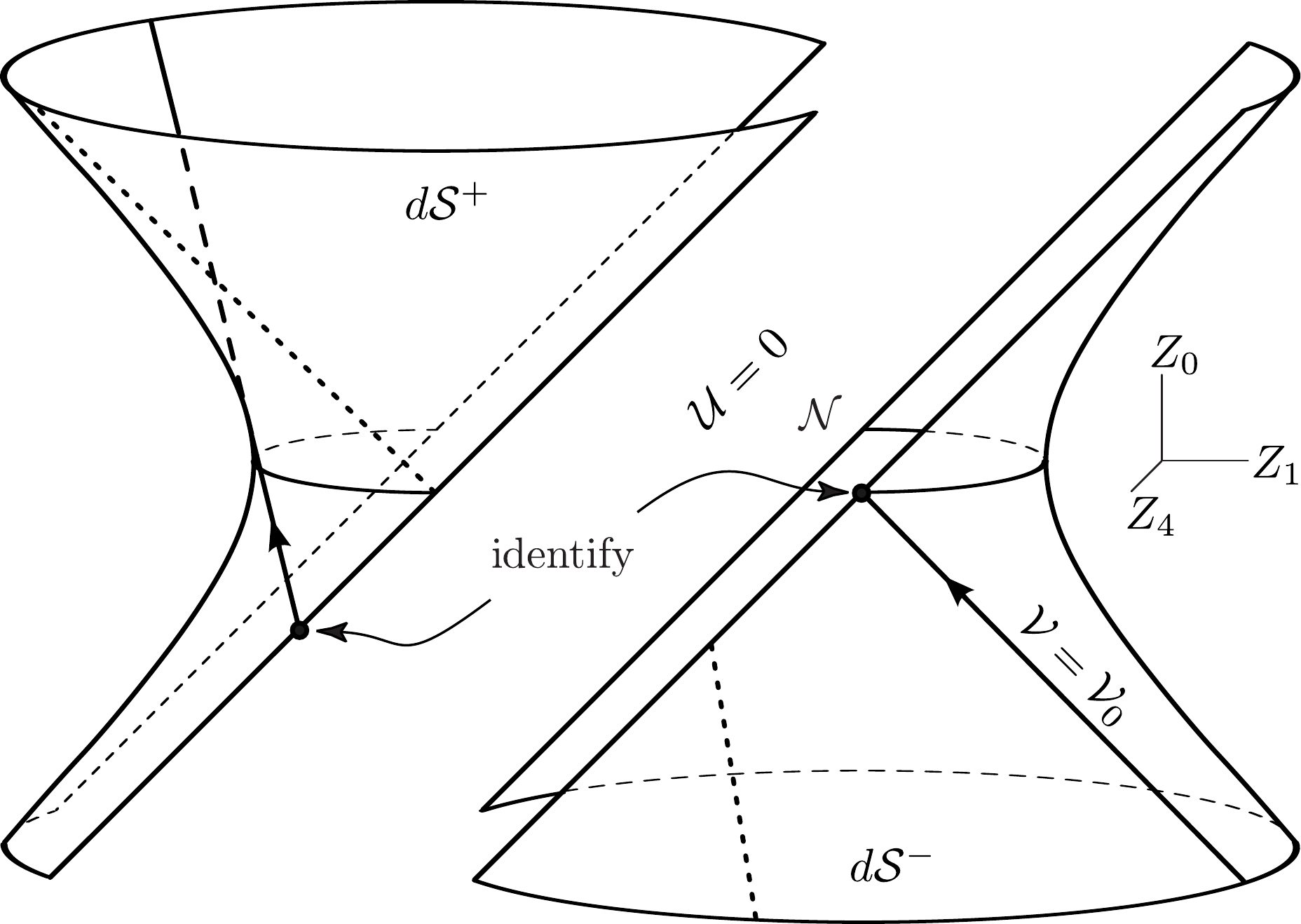}
\caption{The ``cut-and-paste'' construction method for non-expanding impulsive waves in de~Sitter space \eqref{backgr} is based on cutting the hyperboloid \eqref{Constraint_Hyp} along the null hyperplane~$\mathcal{N}$ given by ${\U=0=U}$. The two ``halves'' $d\mathcal{S}^\pm$ are then re-attached with a shift along $\mathcal{N}$. Here we depict the null geodesic generator of $d\mathcal{S}^-$ given by ${\V=\V_0}$. Instead of continuing as unbroken null generator into $d\mathcal{S}^+$ (indicated by the dashed line in the upper left part), its interaction with the impulse at $\mathcal{N}$ makes it jump according to Penrose's identification of points ${\V \mapsto \V - \h^\imag}$, but also refracts it to \emph{become the appropriate null generator of} $d\mathcal{S}^+$, as exactly described by formulae \eqref{5DNullGeod}--\eqref{5DCoeffSpec}. As in Fig.~\ref{fig4}, the directions $Z_2$ and $Z_3$ are suppressed. To ensure that ${Z_2(\lambda)=0= Z_3(\lambda)}$ for all values of the affine parameter~$\lambda$, here we specialise to initial data ${x_0=0=y_0}$ and to waves with ${\h^\imag_{,2}=0=\h^\imag_{,3}}$, so that ${A_2=0=A_3}$.}
\label{fig5}
\end{figure}
\end{center}

This insight provides us with a vivid geometrical picture illustrating the ``cut-and-paste'' approach to the construction of nonexpanding impulsive waves in a background {\bf de Sitter space}, shown in Fig.~\ref{fig5}.


The de Sitter hyperboloid is cut into two ``halves'' $d\mathcal{S}^-$ and $d\mathcal S^+$ along the non-expanding spherical impulsive surface ${\U=0=U}$. These two parts are then re-attached in the very specific and unique way:
The generators approaching the impulse at ${\U=0}$ from $d{\mathcal S}^-$ are {\bf shifted} from their value ${\V_0}$ in $d{\mathcal S}^-$ to the value ${\V^0+\h^\imag}$ in $d{\mathcal S}^+$ due to the interaction with the wave.\footnote{In terms of the coordinates of the 5-dimensional embedding metric \eqref{5D_imp}, \eqref{Constraint_Hyp-again}, such a shift  reads ${V^0+\tfrac{1}{2}H^\imag }$ because ${V^0=\V_0/\alpha}$ and ${\h^\imag = \tfrac{1}{2}\alpha H^\imag}$, see \eqref{GenPos}, \eqref{H-relations}, \eqref{4DGeodAfBetaNeq0}.} Moreover, they are tilted (refracted) by the impulse according to \eqref{5DNullGeod}, {\bf which is precisely the amount needed to turn them into the generators of} ${d\mathcal S}^+$ starting at ${\V^0+\h^\imag}$.

\newpage

\begin{center}
\begin{figure}[h]%
\hspace{30mm}\includegraphics[scale=0.48]{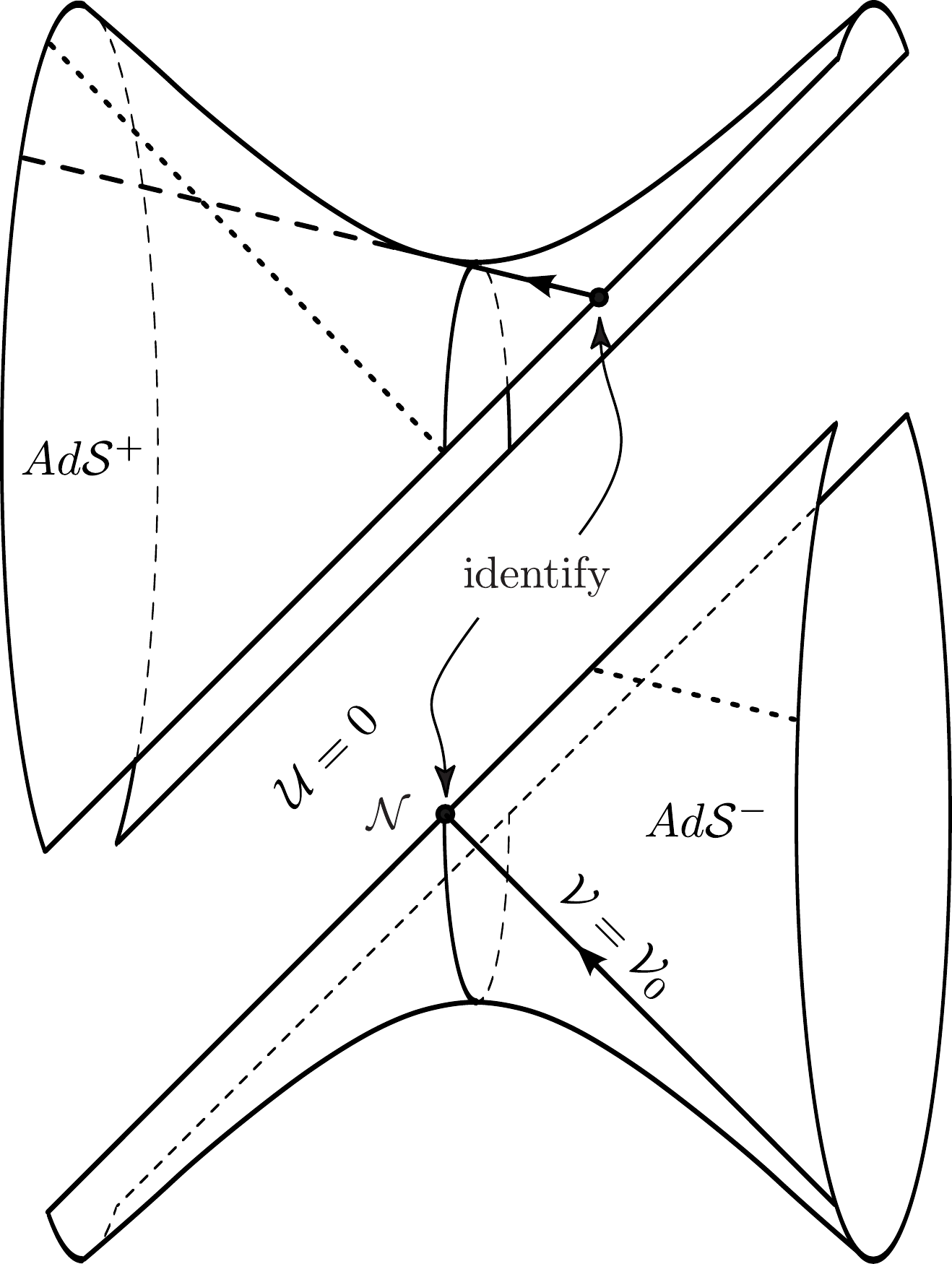}
\caption{The ``cut-and-paste'' construction method for non-expanding impulsive waves in anti-de~Sitter space \eqref{backgr} is based on cutting the hyperboloid \eqref{Constraint_Hyp} along the null hyperplane~$\mathcal{N}$ given by ${\U=0=U}$. The two ``halves'' $Ad\mathcal{S}^\pm$ are then re-attached with a shift along $\mathcal{N}$. Here we depict the null geodesic generator of $Ad\mathcal{S}^-$ given by ${\V=\V_0}$. Instead of continuing as unbroken null generator into $Ad\mathcal{S}^+$, its interaction with the impulse at $\mathcal{N}$ makes it jump according to the Penrose's identification of points ${\V \mapsto \V - \h^\imag}$, but also refracts it to \emph{become the appropriate null generator of} $Ad\mathcal{S}^+$, as exactly given by \eqref{5DNullGeod}--\eqref{5DCoeffSpec}.}
\label{fig6}
\end{figure}
\end{center}

The picture representing the construction of nonexpanding impulsive waves {\bf in anti-de Sitter space is analogous}, starting with the hyperboloid shown on the right of Fig.~\ref{fig6}. Since now ${\Lambda<0}$ the corresponding sign parameter is ${\sigma=-1}$, so that the impulse \eqref{imp_surface} has the geometry of $H^2$. Nevertheless, the transformation \eqref{discontinuous-transformation} and the inherent Penrose junction conditions have the same form. The geometrical meaning of the shift and the refraction on the null generators of the anti-de Sitter hyperboloid thus remain the same. The corresponding picture is presented in Fig.~\ref{fig6} .

\vspace*{-2mm}

We may thus conclude that these explicit visualizations provide us with clear geometrical insights. They give a deeper understanding of the various construction methods of nonexpanding impulsive gravitational waves propagating in de Sitter and anti-de Sitter universes. Moreover, they naturally explain their unambiguous mutual relations, because the key elements of the geometric picture --- namely the null generators of the hyperboloids representing the constant-curvature backgrounds --- are globally unique.

\newpage

\backmatter

\bmhead{Acknowledgments}

This paper was supported by the Czech Science Foundation Grant No.~GA\v{C}R 20-05421S (JP) and the Austrian Science Fund FWF-Grant No.~P33594 (RS). The authors thank their frequent co-authors Clemens S\"amann and Robert \v{S}varc, who were involved in much of the works summarized here. Finally, we are very grateful to Roger Penrose for inspiring us through his work for all those years.
\vspace{-3mm}

\begin{center}
\begin{figure}[ht]
   \centering
   \includegraphics[scale=0.9]{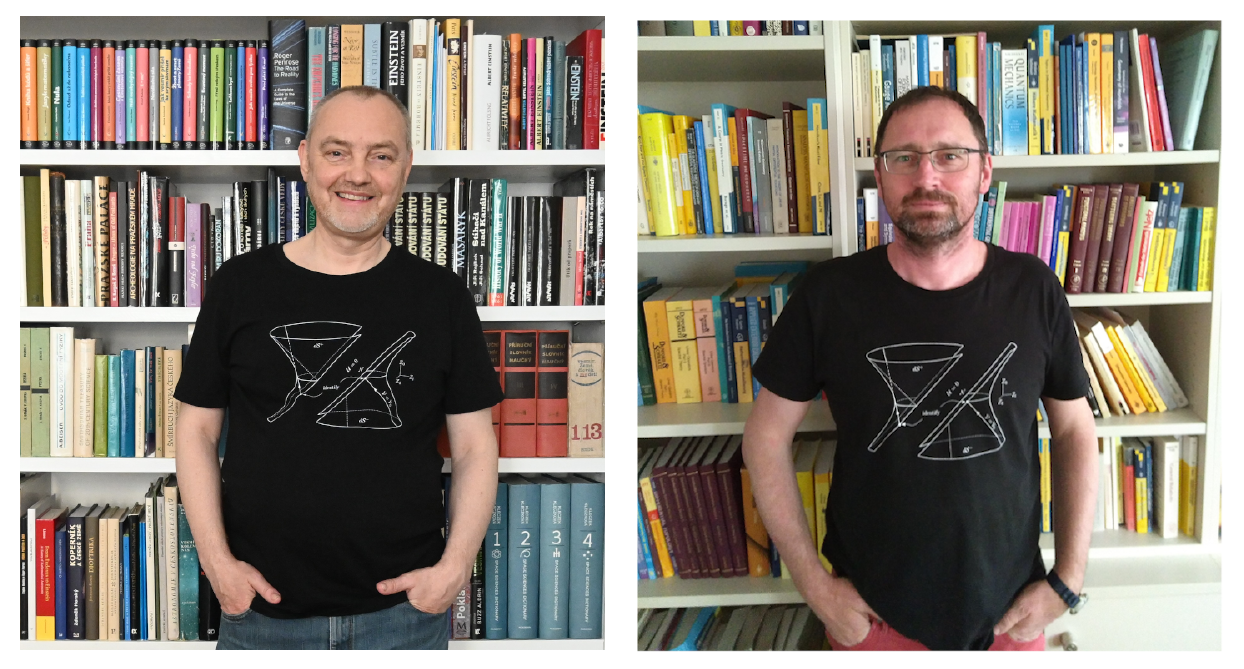} 
\end{figure}
\end{center}
\
\vspace{-16mm}

\section*{Declarations}
%
%
\begin{itemize}

\item Availability of data and materials

Not applicable to this article because it is based on purely theoretical considerations, without using any datasets or other materials.


 \item Authors' contributions

Both authors contributed equally to this work.

\end{itemize}
%
%




\end{document}